# GIS-based support for the complex botanical studies at the Molnieboi Spur, Altai

I.V. Florinsky[1*], E.V. Selezneva[1], A.I. Kulikova[2]

[1] Institute of Mathematical Problems of Biology, Russian Academy of Sciences
Pushchino, Moscow Region, 142290, Russia

[2] Central Siberian Botanical Garden, Siberian Branch, Russian Academy of Sciences
101, Zolotodolinskaya St., Novosibirsk, 630090, Russia

## Abstract

The Molnieboi Spur is located at the northwestern margin of the Katun Range, the high-mountain part of the Altai Mountains. Unique geological and geophysical characteristics of the Molnieboi Spur made it an attractive target for complex botanical studies including botanical, soil, geological, geochemical, geophysical, radiation, and soil gas surveys and analyses. In this paper, we present the first version of the geographic information system (GIS) application for the Molnieboi Spur developed using the software QGIS. A digital elevation model for the study area was derived from a detailed topographic map. The database was filled with tabular data on about 100 parameters including: eight botanical characteristics of the *Lonicera caerulea* local population, two cytogenetic indices of *Lonicera caerulea* seeds, five types of biochemical parameters of *Lonicera caerulea* leaves and fruits, three types of geochemical characteristics of the local soils, three types of radiation parameters of the local soils and *Lonicera caerulea* plants, and one soil gas parameter. The results of the magnetometric survey were inserted as a raster image. A visual analysis of the maps produced allows one to better understand the spatial relationships between various natural components of the Molnieboi Spur.

## 1. Introduction

The Molnieboi Spur (Fig. 1) is located at the northwestern margin of the Katun Range, the high-mountain part of Altai (Fig. 2a), approximately 5 km south of the Village of Verkh-Uimon (the Ust-Koksa Region, Altai Republic, Russia). The spur has a length of about 2.4 km, a width of up to 100 m, and a relative height of up to 500 m above the adjacent Okol depression-graben (Boyarskikh et al., 2015b).

The Molnieboi Spur is composed from a rock plate of the epidote-amphibolite facies of metamorphism including alternating green (quartz-chlorite) schists, amphibolites, granites, and granite-gneisses. There are solitary plagioclase porphyry dykes and skarn zones. Green schists have undergone metasomatic alteration and include ore minerals, such as hematite and magnetite. All rocks are highly fragmented and fractured (Boyarskikh et al., 2015b).

The spur has a stepped shape caused by a system of reactivated ancient and modern local faults. Zones of these faults are topographically manifested as scarps on the crest of the spur, and gullies on its slopes (Boyarskikh et al., 2015b).

"Molnieboi" is Russian for "lightning strike", because the spur attracts lightnings although it is lower than the surrounding mountains (Fig. 1). Moreover, large ball lightnings have been repeatedly observed over the spur (Dmitriev et al., 2011). Magnetometric measure-ments have revealed a very uneven spatial distribution of the magnetic field at the Molnieboi Spur (Dmitriev et al., 1989; Dmitriev, 1998; Shitov, 1999, 2005; Bakiyanov and Gvozdarev, 2013; Boyarskikh et al., 2012, 2015b). There are at least three pronounced magnetic anomalies with the size of about $20 \times 50$ m. Each of them has a dipole-like structure with a positive and negative sections (up to 12 µT and –6 µT, correspondingly). In general, the positive anomaly is spatially correlated with outcrops containing magnetite (Boyarskikh et al., 2015b). A natural manifestation of the Gibbs phenomenon (Jerri, 1998) may probably explain the origin of these micro-anomalies.

---

* Correspondence to: iflorinsky@yahoo.ca

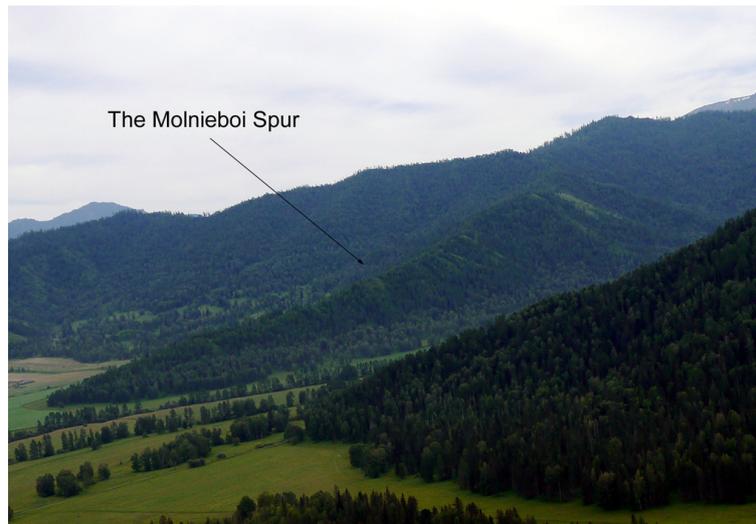

Fig. 1. The Molnieboi Spur: bird's-eye view from the northwest.

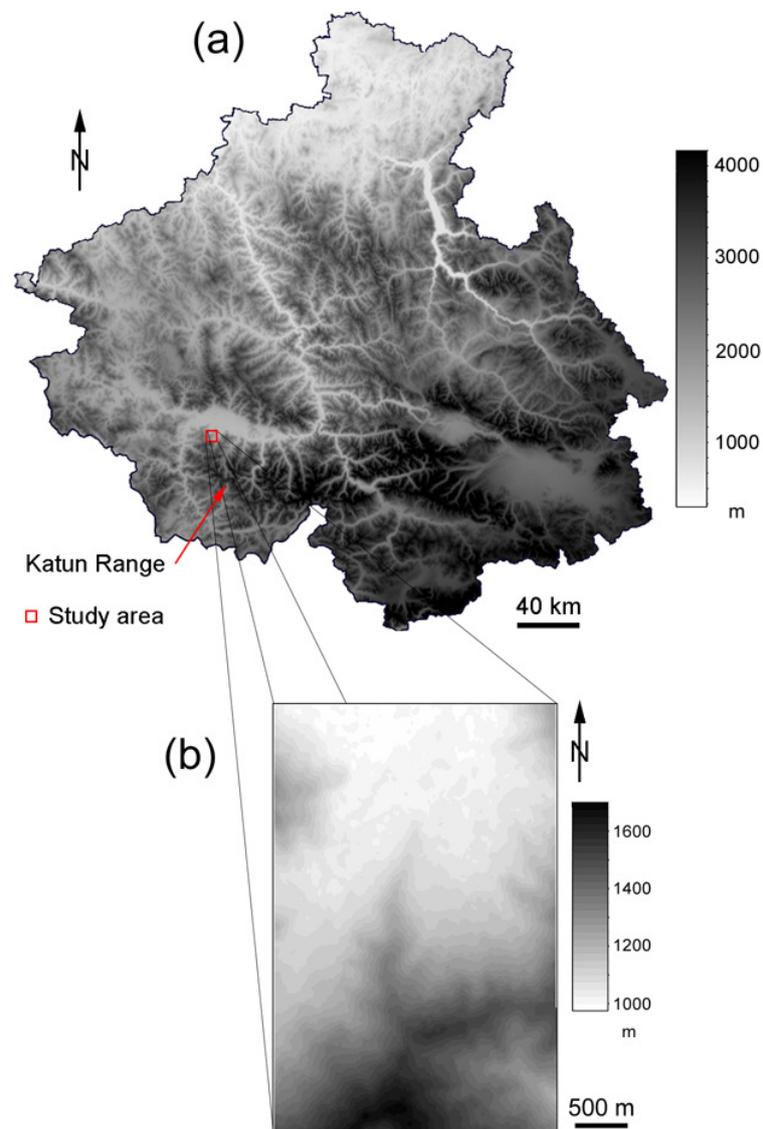

Fig. 2. Location of the Molnieboi Spur (50°10'30" N, 85°42'30" E), elevation maps: (a) regional scale (the Altai Mountains) derived from the global DEM SRTM30_Plus (Sandwell et al., 2008); and (b) local scale (the spur and adjacent area) derived from the ASTER GDEM.

Such unique geological and geophysical characteristics of the Molnieboi Spur made it an attractive target for geophysical (Dmitriev et al., 1989; Dmitriev, 1998; Shitov, 1999, 2005; Dmitriev et al., 2011; Bakiyanov and Gvozdarev, 2013), physiological (Dmitriev and Shitov, 2000), and cytological (Dmitriev et al., 2013) research as well as complex botanical studies including botanical, soil, geological, geochemical, geophysical, radon, and radiation surveys and related laboratory analyses (Boyarskikh et al., 2012, 2013, 2014, 2015a, 2015b, 2015c; Kulikova and Boyarskikh, 2014, 2015).

In the case of the complex botanical studies, various responses of a *Lonicera caerulea* (blue honeysuckle) natural population on geoenvironmental conditions were examined. For this purpose, a study area with the size of about $150 \times 100$ m was previously chosen on the crest and adjacent slopes of the spur. Within the study area, five study sites for detailed botanical, soil, geological, geochemical, geophysical, radiation, and soil gas surveys were previously selected; the study sites have sizes ranging from 300 to 600 m$^2$ (Boyarskikh et al., 2012, 2015b).

It is obvious that modeling and combined analysis of spatially distributed heterogeneous data should be carried out with geo-information technologies. Surprisingly, although the Molnieboi Spur has been studied for more than 35 years, hitherto there were no attempts to map, model, and study this object using both traditional cartographic approaches and geographic information systems (GIS). In this paper, we present the first version of the GIS application for the Molnieboi Spur developed with the software QGIS.

## 2. Materials and methods
### 2.1. GIS software

We utilized QGIS 2.2 to develop the GIS application for the Molnieboi Spur. QGIS is a cross-platform, free and open-source desktop GIS (QGIS, 2009–2015; Westra, 2014; Menke, 2015). QGIS is the international project of the Open Source Geospatial Foundation. It is distributed for free under the GNU General Public License. QGIS supports most raster and vector formats as well as the spatial databases. QGIS has a user-friendly interface and allows performing all the basic operations to analyze and visualize spatial data.

### 2.2. Digital terrain modeling

It is obvious that a digital elevation model (DEM) should be used as a basis for such a GIS application. To select the appropriate DEM, we evaluated two free, publicly available high-resolution DEMs: ASTER GDEM and SRTM1 DEM (USGS, 2015).

ASTER GDEM has a resolution of $1" \times 1"$ that is approximately $30.90$ m $\times 19.85$ m at the latitude of the Molnieboi Spur (50º10'30" N). From one of the ASTER GDEM tiles, we extracted a portion (the matrix $142 \times 140$) with the spur and adjacent area. Then, applying the Delaunay triangulation and a piecewise quadric polynomial interpolation (Agishtein and Migdal, 1991), we produced a square-gridded DEM with the grid size of 9 m (Fig. 2b). To suppress the high-frequency noise, the DEM was smoothed three times. Finally, digital models of several morphometric variables were derived from the smoothed DEM (Fig. 3), namely: slope gradient, slope aspect, horizontal curvature, vertical curvature, minimum curva-ture, maximum curvature, catchment area, and dispersive area. Definitions and physical interpretations of the morphometric attributes as well as description of derivation methods can be found elsewhere (Florinsky, 2012). Data processing was carried out by the software LandLord (Florinsky, 2012, pp. 315–316).

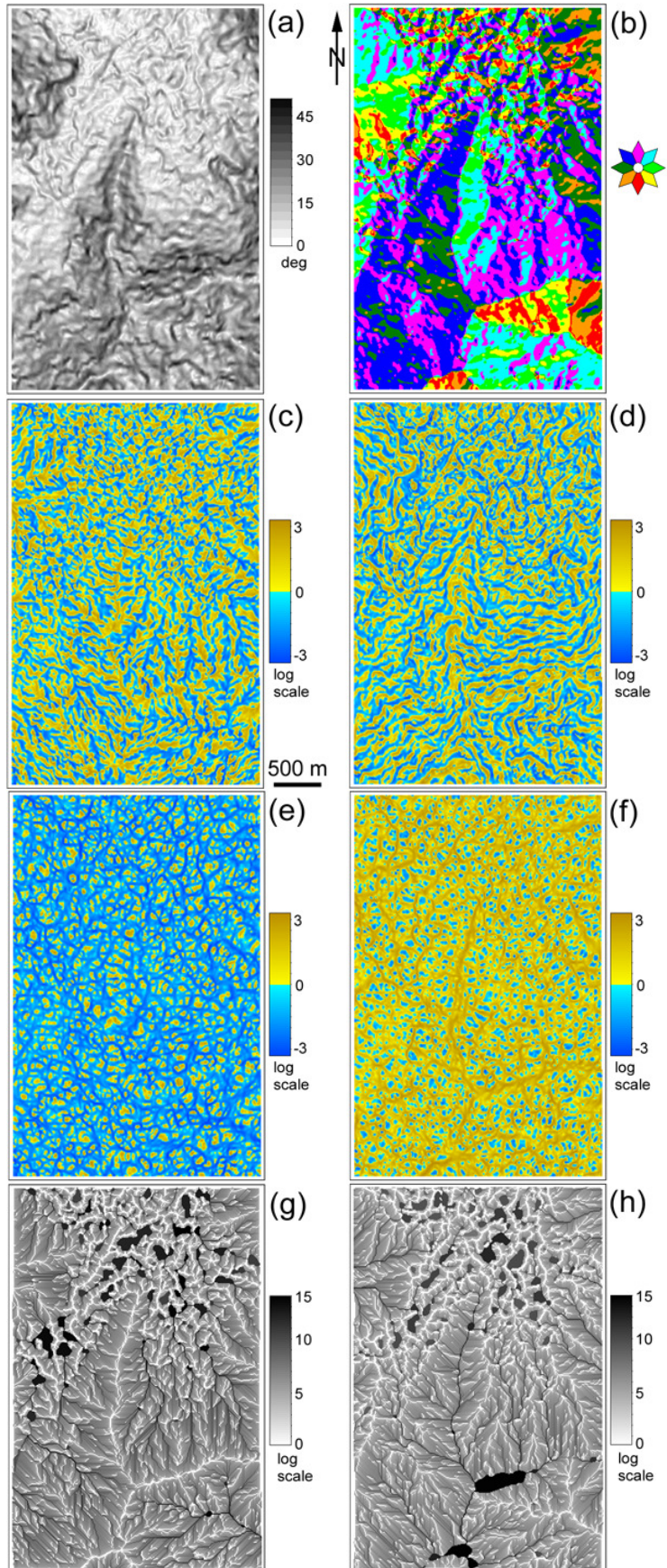

Fig. 3. The Molnieboi Spur, morphometric variables derived from the ASTER GDEM (Fig. 2b): (a) slope gradient; (b) slope aspect; (c) horizontal curvature; (d) vertical curvature; (e) minimum curvature; (f) maximum curvature; (g) catchment area; and (h)dispersive area.

The ASTER-derived digital terrain models (DTMs) with the resolution of 9 m (Figs. 2b and 3) can be used to study the Molnieboi Spur at the local scale. However, such a resolution is insufficient for adequate representation and modeling of the study area, which has the size of about 150 × 100 m. Notice that for territories located to the north of 50º N, SRTM1 DEM has the lower resolution than ASTER GDEM, that is, 1" × 2". So, SRTM1 DEM cannot be also utilized to represent adequately the study area.

Thus, to generate a DEM for the study area, we decided to use a detailed topographic map. Contours of the map were digitized using QGIS. Then, the DEM with the resolution of 2 m was created using the interpolation module of QGIS (Fig. 4a). We also derived digital models of several morphometric variables from this DEM, namely: hill shading (Fig. 4b), slope gradient (Fig. 4c), and slope aspect (Fig. 4d).

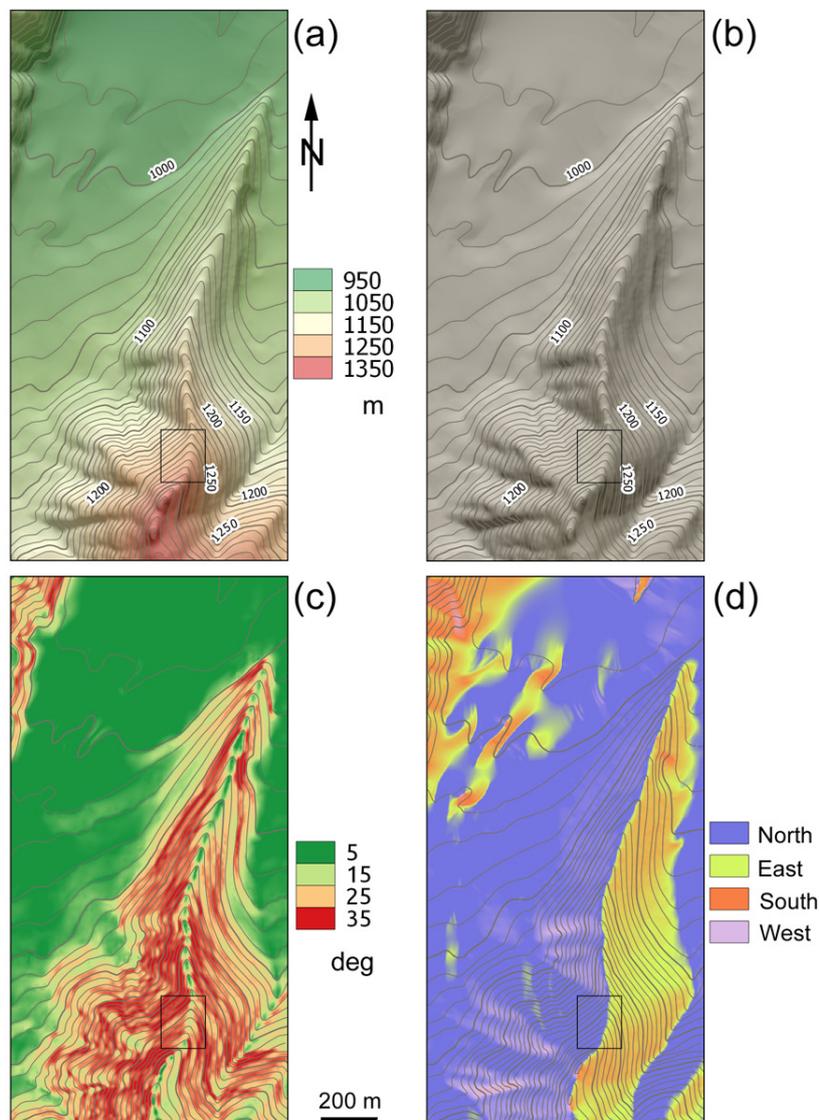

Fig. 4. The Molnieboi Spur, morphometric variables derived from the detailed topographic map: (a) elevation; (b) hill shading; (c) slope gradient; and (d) slope aspect. The rectangle marks the study area.

### *2.3. Database preparation*

To fill the database, we used a data set from botanical, soil, geological, geochemical, geophysical, radiation, and soil gas surveys and related laboratory analyses (Table 1) including about 100 parameters, such as: eight botanical characteristics of the *Lonicera caerulea* local population (Kulikova and Boyarskikh, 2015; Kulikova, 2015), two cytogenetic indices of *Lonicera caerulea* seeds (Boyarskikh et al., 2015b), five types of biochemical characteristics of *Lonicera caerulea* leaves and fruits (Boyarskikh et al., 2013, 2014, 2015a, 2015c), three types of geochemical characteristics of the local soils (Boyarskikh et al., 2013, 2015a, 2015c), three types of radiation parameters of the *Lonicera caerulea* local plants and soils (Boyarskikh et al., 2015c), and one soil gas parameter (Boyarskikh et al., 2015c).

All data were initially presented as Excel tables. Each record has a unique number (or name), coordinates in WGS84 obtained during the fieldwork with a GPS receiver, and attribute information. All data can be divided into two categories: point data relating to individual honeysuckle plants, and polygonal data describing generically five sites within the study area.

Tabular data were geocoded in QGIS and saved as shape files. All data were grouped on the semantic basis (e.g. botany, geochemistry, etc.) as well as according to the type of geometry (point or polygon). The final dataset consists of sixteen vector layers including ten polygonal layers describing five sites within the study area, and six point layers describing individual plants.

The results of the magnetometric survey were presented as a map in TIFF format (Boyarskikh et al., 2015b). It was geocoded using QGIS and saved as a raster layer.

### *2.4. Visualization*

To visualize point data, we applied graduated colored symbols (Figs. 5a and 6). Color tinting was utilized to visualize the raster magnetometric map (Fig. 5b). To visualize polygonal data, we used graduated shading (Fig. 7).

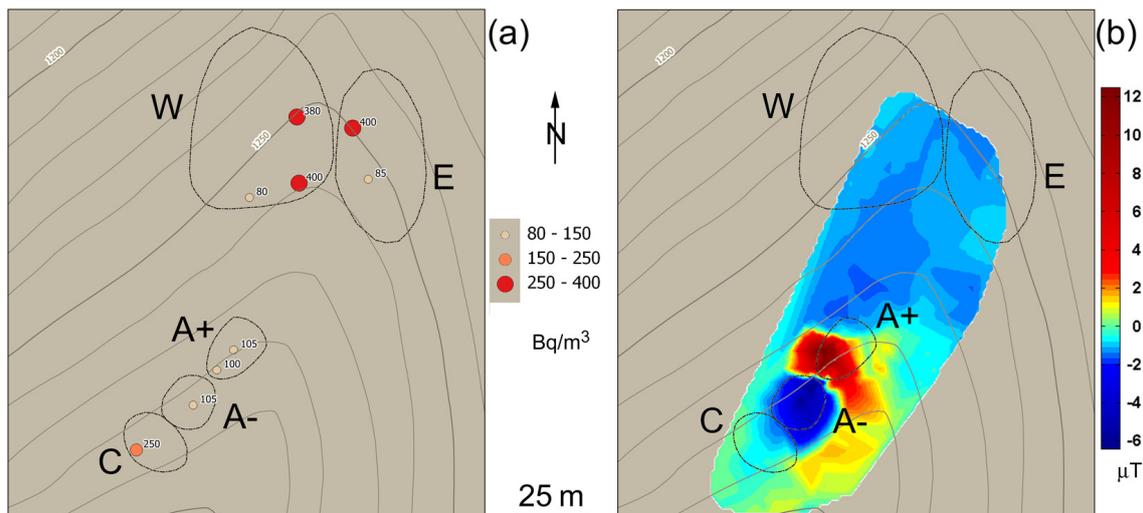

Fig. 5. Some geochemical and geophysical characteristics of the study area: (a) radon content in the soil air; and (b) anomalous magnetic field, the normal magnetic field is 60 μT (modified from Boyarskikh et al., 2015b). A+, A−, W, E, and C are the study sites.

Table 1. Botanical, cytogenetic, biochemical, geochemical, radiation, and soil gas parameters of the study area inputted into the database of the GIS application.

| Data type | Parameter, unit (year of sample collection and/or measurement) |
| --- | --- |
| Botanical | 1. Weight of fruits, g (2012).<br>2. Average number of mature seeds per fruit, % (2012 and 2014).<br>3. Average width of the pollen grains, μm (2014).<br>4. Average length of the pollen grains, μm (2014).<br>5. Germination capacity of seeds, % (2014).<br>6. Energy of germination, % (2014).<br>7. Embryolethality, % (2012, 2014).<br>8. Fertility of pollen grains, % (2014). |
| Cytogenetic | 1. Mitotic index, % (2012).<br>2. Proportion of abnormal mitoses, % (2012). |
| Biochemical | 1. Content of the elements in the leaves, mg/kg (2011–2012) (synchrotron X-ray fluorescence spectrometry):<br> • K, Ca, Fe, Mn, Zn, Cu, Br, Rb, Sr, and Pb.<br>2. Content of the elements in the leaves, mg/kg (2011–2012) (absorption spectroscopy):<br> • K, Ca, Fe, Mn, Cu, Mg, Ni, Li, Sr, and Pb.<br>3. Content of the elements in the fruits, mg/kg (2011–2012) (absorption spectroscopy):<br> • K, Ca, Fe, Mn, Zn, Cu, Mg, Ni, Li, Na, Sr, and Pb.<br>4. Content of biologically active substances in the leaves, % (2011–2012):<br> • Hydroxycinnamic acids;<br> • Flavonoles;<br> • Flavones.<br>5. Content of biologically active substances in the fruits, % (2011–2012):<br> • Hydroxycinnamic acids;<br> • Anthocyans;<br> • Flavonoles;<br> • Flavones. |
| Geochemical | 1. Content of major oxides in the soil, % (2011–2012):<br> • $SiO_2$, $TiO_2$, $Fe_2O_3$, FeO, MnO, MgO, CaO, $Na_2O$, $P_2O_5$, and $K_2O$.<br>2. Content of mobile forms of the elements in the soil, mg/kg (2011–2012) (absorption spectroscopy):<br> • K, Ca, Fe, Mn, Zn, Cu, Na, Mg, Ni, Li, and Sr.<br>3. Total content of the elements in the soil, mg/kg (2011–2012) (synchrotron X-ray fluorescence spectrometry):<br> • K, Ca, Fe, Mn, Zn, Cu, Br, Rb, and Sr. |
| Radiation | 1. Radionuclide activity in the shoots, Bq/kg (2012):<br> • $K^{40}$, $Pb^{210}$, $Pb^{212}$, $Pb^{214}$, and $Be^7$.<br>2. Radionuclide activity in the soil, Bq/kg (2009–2010):<br> • $K^{40}$, $Cs^{137}$, $Ra^{226}$, $Be^7$, and $Pa^{234}$.<br>3. Radionuclide activity in the soil, Bq/kg (2012):<br> • $K^{40}$, $Cs^{137}$, $Pb^{210}$, $Pb^{212}$, and $Pb^{214}$. |
| Soil gas | 1. Radon content in the soil air, $Bq/m^3$ (2014). |

Botanical, cytogenetic, and biochemical characteristics as well as the first radiation one relate to the local *Lonicera caerulea* population.

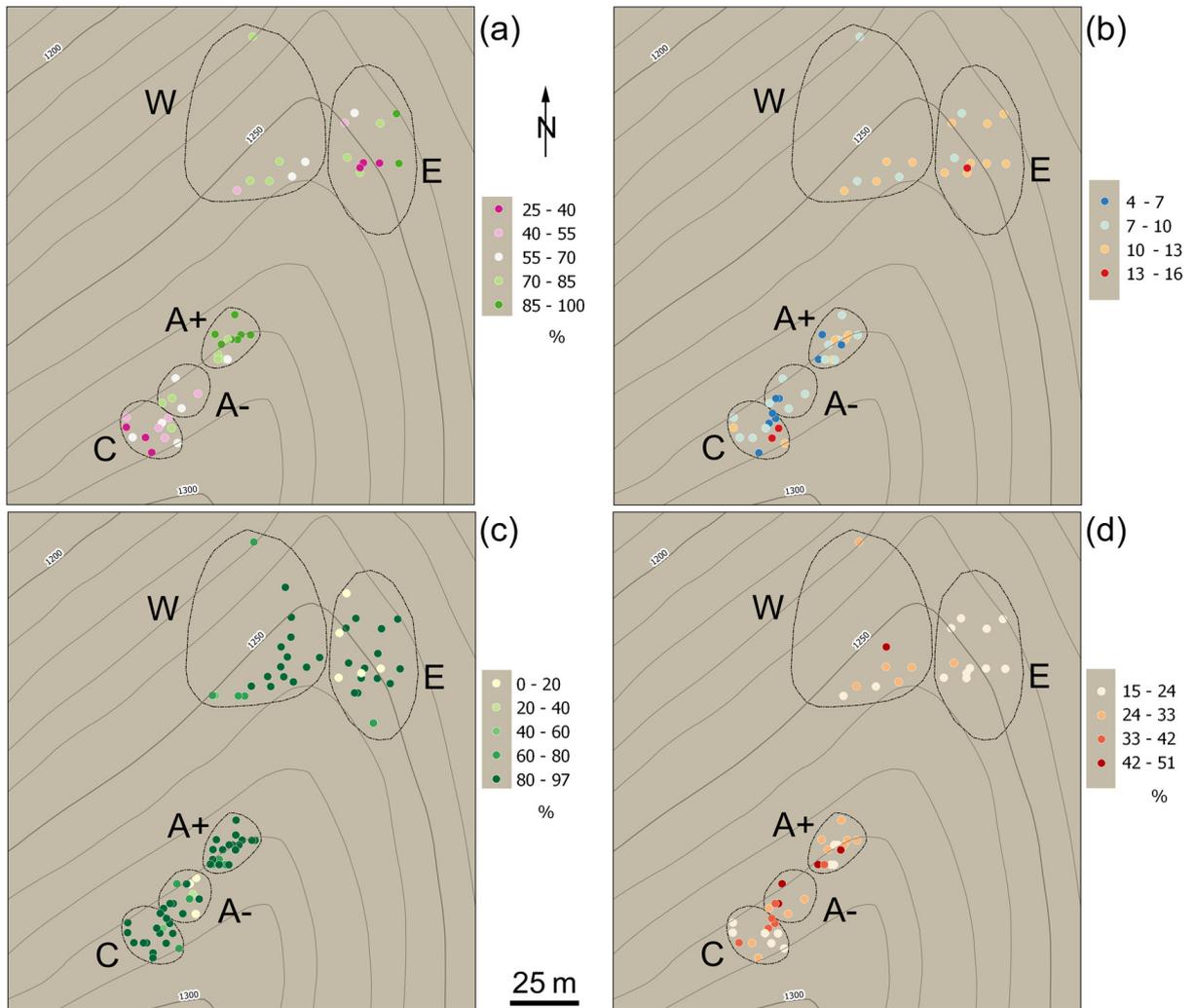

Fig. 6. Some botanical characteristics for individual *Lonicera caerulea* plants of the study area: (a) germination capacity of seeds; (b) average number of mature seeds per fruit; (c) fertility of pollen grains; and (d) embryolethality. A+, A−, W, E, and C are the study sites.

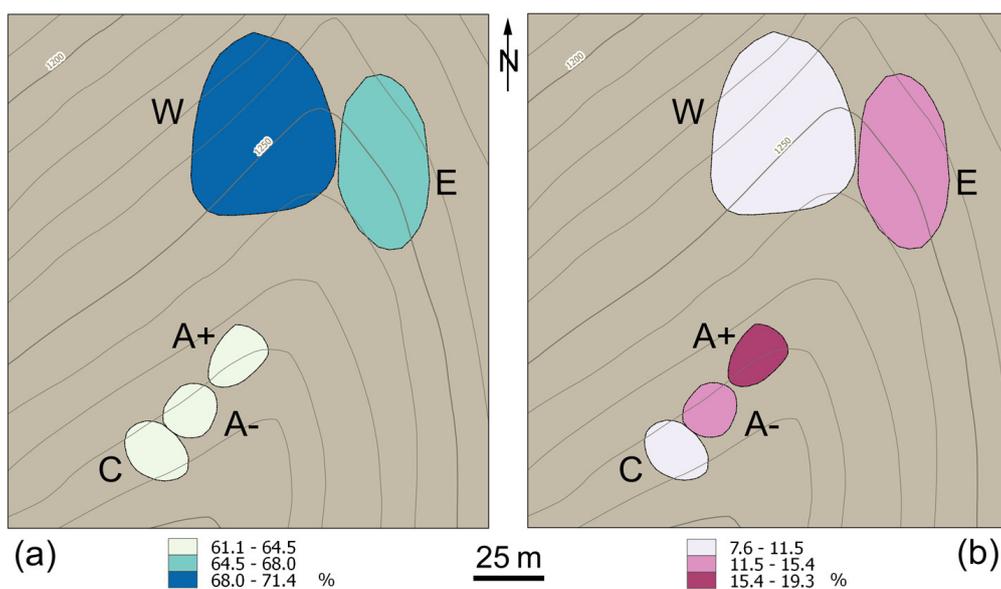

Fig. 7. Some cytogenetic characteristics for the *Lonicera caerulea* population of the study area: (a) mitotic index; and (b) proportion of abnormal mitoses. A+, A−, W, E, and C are the study sites.

### 3. Results

Figures 5–7 represent examples of maps produced with the GIS application developed.

Figure 5 demonstrates some maps of geochemical and geophysical characteristics of the study area. For example, the increased values of the radon content in the soil air reveal clearly a local geological fault crossing the study sites W and E (Fig. 5a), as was supposed by Boyarskikh et al. (2015c). The structure of one of the dipole-like magnetic anomalies is well seen in the southern part of the study area (Fig. 5b) (see details elsewhere – Boyarskikh et al., 2015b).

Figure 6 shows several maps of botanical characteristics for individual *Lonicera caerulea* plants within the study area: germination capacity of seeds, average number of mature seeds per fruit, fertility of pollen grains, and embryolethality. The detailed discussion on relationships between these parameters and geoenvironmental factors can be found elsewhere (Kulikova and Boyarskikh, 2015).

Figure 7 displays two maps of cytogenetic characteristics of the *Lonicera caerulea* population averaged for the study sites, such as mitotic index and proportion of abnormal mitoses. Boyarskikh et al. (2015b) presented a comprehensive discussion on the dependence of cytogenetic parameters on geoenvironmental conditions of the study area.

### 4. Conclusions

We developed the first GIS application for the Molnieboi Spur, the intriguing natural object of the Altai Mountains. The GIS application is intended for the geoinformation support of complex botanical studies at the spur. Mapping can be carried out using various combinations of the parameters from the database. A visual analysis of the maps produced allows one to better understand the spatial relationships between various natural components of the spur. The GIS application can be used as the basis for the further filling with future results of geological and geophysical surveys as well as other spatially distributed data collected at the spur.

To improve the topographic layers of the GIS application, one may wish to derive a DEM of the spur with the resolution of 0.5 m or higher. This is possible using a tacheometric survey or an aerial LiDAR survey from a helicopter or drone.

**Acknowledgements**

The study was supported by the RFBR grant 15-37-50800. The authors are grateful to I.G. Boyarskikh (Central Siberian Botanical Garden) for the organization of field surveys and data, A.R. Agatova (Sobolev Institute of Geology and Mineralogy) for the geological survey and data, A.S. Syso and S.A. Khudyaev (Institute of Soil Science and Agrochemistry) for the soil and geochemical surveys and data, as well as A.I. Bakiyanov (Gorno-Altaisk State University) for the magnetometric survey and data as well as the GPS survey of plants.

### References

Agishtein, M.E., and Migdal, A.A., 1991. Smooth surface reconstruction from scattered data points. *Computers and Graphics*, 15: 29–39.

Bakiyanov, A.I., and Gvozdarev, A.Y., 2013. Geophysical study of the Molnieboi Spur. *Junior Scientists Bulletin of the Gorno-Altaisk State University*, No. 10: 179–185 (in Russian).

Boyarskikh, I.G., Chankina, O.V., Khudyaev, S.A., and Syso, A.I., 2013. Investigating the elemental composition of a soil-plant system, based on the example of *Lonicera caerulea. Bulletin of the Russian Academy of Sciences: Physics*, 77: 191–194.

Boyarskikh, I.G., Chankina, O.V., Syso, A.I., and Vasiliev, V.G., 2015a. Trends in the content of chemical elements in leaves of *Lonicera caerulea* (Caprifoliaceae) in connection with their secondary metabolism in the natural populations of the Altai Mountains. *Bulletin of the Russian Academy of Sciences: Physics*, 79: 94–97.

Boyarskikh, I.G., Kulikova, A.I., Agatova, A.R., Bakiyanov, A.I., and Florinsky, I.V., 2015b. Mutation activity of *Lonicera caerulea* population in an active fault zone (the Altai Mountains).

*arXiv*:1508.02016, 12 p.

Boyarskikh, I.G., Syso, A.I., Khudyaev, S.A., Bakiyanov, A.I., Kolotukhin, S.P., Vasiliev, V.G., and Chankina, O.V., 2012. Specific features of elemental and biochemical composition of *Lonicera caerulea* L. in local geologically active zone of the Katun Range (Altai Mountains). *Geophysical Processes and Biosphere*, 11(3): 70–84 (in Russian, with English abstract).

Boyarskikh, I.G., Syso, A.I., and Mazhejka, J., 2015c. Geophysical and soil-geochemical characteristics of environment in relation to plant metabolism in a local zone of seismotectonic manifestations of Gorny Altai. In: Puzanov, A.V., Bezmaternykh, D.M., Rozhdestvenskaya, T.A., Baboshkina, S.V., and Troshkin, D.N. (Eds.), *Biogeochemistry of Technogenesis and Current Problems of Geochemical Ecology, Vol. 1: Proc 9<sup>th</sup> Int. Biogeochem. Workshop, 24–28 Aug. 2015, Barnaul*. Institute of Water and Ecological Problems, Barnaul, pp. 73–76 (in Russian).

Boyarskikh, I.G., Vasiliev, V.G., and Kukushkina, T.A., 2014. The content of flavonoids and hydroxycinnamic acids in the leaves and fruits of *Lonicera caerulea* (Caprifoliaceae) from the populations of the Altai Mountains. *Plant Resources*, 50: 105–121 (in Russian, with English abstract).

Dmitriev, A.N., 1998. *Natural Self-Luminous Formations*. Institute of Mathematics, Novosibirsk, 243 p. (in Russian).

Dmitriev, A.N., and Shitov, A.V., 2000. Psychophysiological interactions of operators with geomagnetic field at abnormal sites. *Bulletin of the International Scientific Research Institute of Cosmic Anthropoecology*, 7: 73–81 (in Russian).

Dmitriev, A.N., Gvozdarev, A.Y., Matushkin, Y.G., Voronkov, E.G., Vlasov, V.N., Krechetova, S.Y., Bakiyanov, A.I., Betev, A.A., Belikova, M.Y., Voronkova, A.E., and Velilyaeva, E.S., 2013. *Complex Scientific Study of the Bashadar Mounds Area, 2004–2011*. Gorno-Altaisk State University, Gorno-Altaisk, 120 p. (in Russian).

Dmitriev, A.N., Krechetova, S.Y., and Kocheeva, N.A., 2011. *Thunderstorms and Thunderstorm Forest Fires in the Territory of the Altai Republic*. Gorno-Altaisk State University, Gorno-Altaisk, 154 p. (in Russian).

Dmitriev, A.N., Novikov, G.N., and Skavinsky, V.P., 1989. *Local Geophysical and Geochemical Studies of Tectonophysical Regions of the Altai Mountains*. Institute of Geology and Geophysics, Novosibirsk, 41 p. (in Russian).

Florinsky, I.V., 2012. *Digital Terrain Analysis in Soil Science and Geology*. Academic Press, Amsterdam, 379 p.

Jerri, A.J., 1998. *The Gibbs Phenomenon in Fourier Analysis, Splines and Wavelet Approximations*. Kluwer, Boston, 336 p.

Kulikova, A.I., 2015. Reaction of *Lonicera caerulea* reproductive structures on changes of geoenvironmental characteristics in local areas of seismotectonical manifestations. In: *Abstracts, 3<sup>rd</sup> (11<sup>th</sup>) Int. Botanical Conf. Young Scientists, 4–9 Oct. 2015, St. Petersburg*. Komarov Botanical Institute, St. Petersburg, p. 94 (in Russian).

Kulikova, A.I., and Boyarskikh, I.G., 2014. Peculiarities of reproductive structures formation in the abnormal form of *Lonicera caerulea* subsp. *altaica* (Caprifoliaceae). *Botanical Journal (Moscow)*, 99: 193–205 (in Russian, with English abstract).

Kulikova, A.I., and Boyarskikh, I.G., 2015. Reproductive ability of *Lonicera caerulea* (Caprifoliaceae) in the local area of geological and geophysical heterogeneity in the Altai Mountains. *Contemporary Problems of Ecology*, 8: 503–511.

Menke, K., 2015. *Mastering QGIS*. Packt Publishing, Birmingham, 388 p.

QGIS, 2009–2015. *QGIS*, http://qgis.org

Sandwell, D.T., Smith, W.H.F., and Becker, J.J., 2008. *SRTM30_PLUS V11*. Scripps Institution of Oceanography, University of California, San Diego, ftp://topex.ucsd.edu/pub/srtm30_plus/

Shitov, A.V., 1999. *Natural Self-Luminous Formations as an Eco-Geological Factor of the Mountain Altai Territory*. Ph.D. Thesis. Gorno-Altaisk University, Gorno-Altaisk, 170 p. (in Russian).

Shitov, A.V., 2005. Magnetic characteristics of geological microobjects. *Natural Resources of the Mountain Altai*, No. 4(2): 108–114 (in Russian).

USGS, 2015. *Earth Explorer*. U.S. Geological Survey, Sioux Falls, http://earthexplorer.usgs.gov

Westra, E., 2014. *Building Mapping Applications with QGIS*. Packt Publishing, Birmingham, 248 p.